\documentclass[12pt,letterpaper]{article}
\pdfoutput=1
\usepackage{graphicx,array}
\usepackage{color}
\usepackage{latexsym}
\usepackage{amsthm}
\usepackage{amsmath}
\usepackage{enumitem}
\usepackage{amssymb}
\usepackage{hyperref}
\usepackage[hang,flushmargin]{footmisc}

\def\simgt{\mathrel{\lower2.5pt\vbox{\lineskip=0pt\baselineskip=0pt
           \hbox{$>$}\hbox{$\sim$}}}}
\def\simlt{\mathrel{\lower2.5pt\vbox{\lineskip=0pt\baselineskip=0pt
           \hbox{$<$}\hbox{$\sim$}}}}

\newcommand{\be}{\begin{equation}}
\newcommand{\ee}{\end{equation}}
\newcommand{\bea}{\begin{eqnarray}}
\newcommand{\eea}{\end{eqnarray}}
\newcommand{\Eq}[1]{Eq.~(\ref{#1})}

\newcommand{\Sec}[1]{Sec.~\ref{#1}}

\newcommand{\Fig}[1]{Fig.~(\ref{#1})}
\newcommand{\App}[1]{App.~\ref{#1}}

\newcommand{\lads}{\ell}

\newcommand{\ket}[1]{\left| #1 \right\rangle}

\definecolor{nicered}{rgb}{0.7,0.1,0.1}
\definecolor{nicegreen}{rgb}{0.1,0.5,0.1}

\begin{document}
\thispagestyle{empty}
\hfill
CALT-TH-2014-139
\vspace{4cm}

\begin{center}
{\LARGE\bf
Disrupting Entanglement of Black Holes
}\\
\bigskip\vspace{1cm}{
{\large Stefan Leichenauer}\let\thefootnote\relax\footnotetext{email: sleichen@theory.caltech.edu} 
} \\[7mm]
 {\it Walter Burke Institute for Theoretical Physics \\ 
 California Institute of Technology, Pasadena, CA 91125}
 \end{center}
\bigskip
\centerline{\large\bf Abstract}

\begin{quote} \small
We study entanglement in thermofield double states of strongly coupled CFTs by analyzing two-sided Reissner-Nordstr\"om solutions in AdS. The central object of study is the mutual information between a pair of regions, one on each asymptotic boundary of the black hole. For large regions the mutual information is positive and for small ones it vanishes; we compute the critical length scale, which goes to infinity for extremal black holes, of the transition. We also generalize the butterfly effect of Shenker and Stanford~\cite{Shenker:2013pqa} to a wide class of charged black holes, showing that mutual information is disrupted upon perturbing the system and waiting for a time of order $\log E/\delta E$ in units of the temperature. We conjecture that the parametric form of this timescale is universal. 
\end{quote}

\newpage

\tableofcontents

\newpage

\section{Introduction }

The connection between geometry and entanglement is exciting and deep. In particular, the recent ER=EPR framework introduced by Maldacena and Susskind~\cite{Maldacena:2013xja} suggests that, in a gravitational theory, we should always associate entanglements with wormholes. As discussed in Ref.~\cite{Maldacena:2013xja}, a classical wormhole requires not only a large amount of entanglement but also a very detailed kind of entanglement. They suggested that a ``quantum" wormhole could be associated to any kind of entanglement, though there is no independent meaning to the quantum wormhole as of yet. Here we discuss certain transitions between situations with a large amount of well-ordered entanglement (complete with classical wormhole) to situations where the total entanglement remains large but becomes unordered, and the classical wormhole begins to develop pathologies. 

The systems we study consist of two identical copies of a large-$N$ conformal field theory, ${\rm CFT}_L \times {\rm CFT}_R$ (called the left and right CFTs, respectively), in a particular entangled state $\ket{\Psi} \in \mathcal{H}_L\otimes \mathcal{H}_R$. The left and right CFTs are completely decoupled, meaning that it is impossible to send signals between the two copies. They only know about each other through their entanglement. We choose to place the system in a thermofield double state with inverse temperature $\beta$ and chemical potential $\phi$ (associated to some conserved global $U(1)$ symmetry):
\be\label{eq-tfdoublestate}
\ket{\Psi} = \frac{1}{\sqrt{Z}} \sum_{n} e^{-\frac{\beta}{2}(E_n- \phi Q_n)}\ket{n}_L\otimes \ket{n}_R.
\ee
In AdS/CFT, these types of states are dual to two-sided eternal Reissner-Nordstr\"om black holes~\cite{Maldacena:2001kr}. The two field theories live on the two asymptotic boundaries, and the Einstein-Rosen bridge connecting them is non-traversible as a necessary consequence of the fact that the two field theories are decoupled. This wormhole is a reflection of the entanglement, and its classical nature shows that the entanglement is highly ordered.

It has long been known that the Bekenstein-Hawking entropy of a black hole, as measured by its area, corresponds to the total entanglement entropy between the two copies of the CFT. In other words, the cross-sectional area of the wormhole is determined by the total amount of entanglement of the two subsystems. The length of the wormhole is naturally associated to another measurement of entropy, the mutual information~\cite{VanRaamsdonk:2009ar,  Maldacena:2013xja, Hartman:2013qma, Andrade:2013rra}. For two disjoint regions $A$ and $B$, the mutual information $I(A,B)$ is given by
\be
I(A,B) = S(A) + S(B) - S(A\cup B),
\ee
where $S(\cdot)$ is the von Neumann entropy of the reduced density matrix associated to the given region, obtained by tracing out everything outside of the region.
In the case of the eternal black hole, we wish to consider the  the mutual information between a region $A$ on the left boundary and a region $B$ on the right boundary. The amount of mutual information between two such regions is related to the length of the wormhole connecting them: if we manipulate the state in a way that disrupts this mutual information, we will see the wormhole geometry grow longer.

The computation of mutual information is made possible by the Ryu-Takayanagi prescription for entanglement entropy~\cite{Ryu:2006bv, Hubeny:2007xt}. To a region $A$ on the left asymptotic boundary we associate a minimal surface in the bulk whose boundary coincides with the boundary of $A$. The area of this surface, divided by $4G$, gives the leading contribution to $S(A)$ in the large-$N$ limit.\footnote{We will only be concerned with the leading behavior of entropy and mutual information in the large-$N$ limit. In this approximation, ``vanishing" mutual information only means that the coefficient of $N^2$ vanishes (i.e., the coefficient of $1/G$). It is possible that subleading terms could be accessed by considering corrections to the Ryu-Takayanagi prescription~\cite{Faulkner:2013ana, Lewkowycz:2013nqa}.}   In all of the states we consider, the minimal surface associated to $A$ lies outside of the black hole horizon. A similar procedure gives $S(B)$. To the union $A\cup B$ there are two candidate minimal surfaces which are both extremal, and we must choose the one that actually has the smaller area. The first candidate is the union of the $A$ and $B$ surfaces. This surface consists of two disjoint pieces, and its area is clearly equal to the sum of the areas of its two parts. Thus if this represents the true minimal surface then the mutual information between $A$ and $B$ will be zero. The second candidate surface stretches through the wormhole and connects the two regions. If this surface has area less than the sum of the two disjoint surface areas, then it is the correct minimal surface and the mutual information will be positive. For simplicity, we will focus on the case where the region $B$ on the right is the same as the region $A$ on the left. That is, if we consider the equivalent situation of two uncoupled CFTs living on the same space, then the regions $A$ and $B$ are identical.

We will consider two deformations of the thermofield double states which disrupt the mutual information while leaving the total entanglement intact. The first one is an old idea that we will make explicit: lowering the temperature to zero. In this case the black hole becomes extremal, and in that limit the wormhole becomes infinitely long while retaining a finite area.\footnote{For black holes with compact horizons, when we lower the temperature we risk crossing the Hawking-Page phase transition to a state without a black hole~\cite{Hawking:1982dh}. While this may be a good example of the loss of a classical wormhole, we are not going to study it here. By having a large enough chemical potential, we can avoid this transition~\cite{Hawking:1999dp}.} Any extremal surface crossing the wormhole will be stretched to infinite area, and so the mutual information between any finite regions on the left and right will necessarily drop to zero. Note that the density operator of, say, ${\rm CFT}_L$, obtained from \Eq{eq-tfdoublestate} by tracing out the states of ${\rm CFT}_R$, approaches a projection operator onto the ``ground states" of the effective Hamiltonian $\hat{H} - \phi \hat{Q}$ in the $\beta\to \infty$ limit. There are a great many of these states, which is why the black hole entropy remains finite. 

In \Sec{sec-tempdep} we analyze in detail the loss of mutual information in the transition to an extremal black hole. The state of \Eq{eq-tfdoublestate} contains length scales determined by $\beta$ and $\phi$. It is not surprising that for a region $A$ on the left whose size is much smaller than these length scales the mutual information between it and its partner region on the right is zero; these regions are too small to notice the entanglement. For larger regions the mutual information is positive. There is a critical linear size $L$ of region $A$ for which the transition happens. We compute this critical size and see that it goes to infinity as the temperature goes to zero.

A second way to disrupt the mutual information, while leaving the temperature finite, is to make use of the butterfly effect of Shenker and Stanford ~\cite{Shenker:2013pqa}. A small perturbation is added to the left field theory, which we model as a shift in the energy density. After an amount of time $t_*$, we will find that the mutual information between the two sides is disrupted. This is interpreted as a manifestation of the butterfly effect familiar from chaotic dynamics: a small change in initial condition leads to a large change at later times. Geometrically, this effect is manifested by a shockwave which travels across the horizon of the black hole. Any probe which crosses the horizon, such as the minimal surfaces used to compute $S(A\cup B)$, will be affected by the shock. In \Sec{sec-butterfly} we generalize the analysis of this effect from the BTZ case studied in Ref.~\cite{Shenker:2013pqa} to a wide class of Reissner-Nordstr\"om black holes. We find the apparently universal behavior
\be
t_* \sim \frac{\beta}{2\pi}\log\frac{E}{\delta E},
\ee
where $E$ is the initial energy and $\delta E$ is the energy of the perturbation. Care must be taken in the near-extremal case. There the energy $E$ we use in this formula is not the total energy of the black hole, but the energy in excess of the extremal black hole with the same charge: $E = E_{\rm tot} - E_{\rm ext}$, which goes to zero in the extremal limit. This suggests that only the degrees of freedom excited above the extremal state participate in the chaotic dynamics. This lines up with the fact that the mutual information between local regions drops to zero in the extremal limit: it is only those same excited degrees of freedom which contribute to the mutual information.

It was noted in Ref.~\cite{Shenker:2013pqa}, in the context of the uncharged BTZ black hole, that a natural smallest choice of $\delta E$ is $E/S$, the average energy per degree of freedom, for which $t_*$ becomes the fast scrambling time
\be
t_{\rm sc} \sim  \frac{\beta}{2\pi} \log S.
\ee
Black holes have been conjectured to be fast scramblers, and the fast scrambling timescale has appeared in numerous places in the study of black holes and quantum circuits~\cite{Hayden:2007cs, Sekino:2008he, Lashkari:2011yi}. In the near-extremal case, this should be modified to $\delta E_{\rm min} \sim E/\Delta S$, where $\Delta S = S - S_{\rm ext}$ is the entropy in excess of the extremal entropy.


\section{The Setup}
\label{sec-setup}

We consider two copies of a field theory on $\mathbb{R} \times \Sigma$ in a thermofield double state, where $\Sigma$ is either a sphere $(k=1)$, plane $(k=0)$, or hyperboloid $(k=-1)$ with line element given by
\be
d\Sigma_{d-1}^2 = \lads^2\left[\frac{d\xi^2}{1+ k \xi^2} + \xi^2 d\Omega_{d-2}^2\right].
\ee
By construction, the density operator in either CFT is given by the grand canonical density operator, $\hat{\rho} = \exp[-\beta(\hat{H} - \phi \hat{Q})]/Z$. The AdS dual of the thermofield double state is a two-sided eternal Reissner-Nordstr\"om black hole with metric
\begin{align}
ds^2 &= -f(r) dt^2 + f(r)^{-1} dr^2 + \frac{r^2}{\lads^2}d\Sigma_{d-1}^2,\\
f(r) &= k - \frac{\mu}{r^{d-2}}+\frac{q^2}{r^{2d-4}} + \frac{r^2}{\lads^2}.
\end{align}
The outer horizon is located at $r=R$, the largest root of $f(R)=0$, and the inverse temperature is  $\beta = 4\pi/f'(R)$. The chemical potential is given by a rescaled version of the electric potential difference between the horizon and infinity, $\phi = -q/R^{d-2}$. The details of the thermodynamics of these solutions is reviewed in \App{sec-RNthermo}, but here we record the results for the energy and entropy densities:
\begin{align}
\epsilon &\equiv \frac{\langle \hat{H} \rangle}{{\rm Vol}(\Sigma)} = \frac{d-1}{16\pi G \lads^{d-1}}(\mu - \mu_0), & s &\equiv \frac{\langle -\log \hat{\rho} \rangle}{{\rm Vol}(\Sigma)} = \frac{1}{4G} \left(\frac{R}{\lads}\right)^{d-1}.
\end{align}
The zero-point energy $\mu_0$ is equal to zero for $k=0,1$, but equal to a finite negative value for $k=-1$. This is because vacuum AdS is not the lowest energy solution when $k=-1$, and we refer to \App{sec-RNthermo} for details.\footnote{Ref.~\cite{Emparan:1999pm} also considered Casimir energy corrections to this equation, which are non-zero for $k=1$ as well. Those energies are independent of the mass and charge, so we could likewise absorb them into $\mu_0$.} As we will see, it is natural in the present context to measure energies relative to the $\mu= \mu_0$ state. Even though we calculate these quantities using holography, they have interpretations purely in the field theory. If we like, we can think of these as the defining equations for the gravitational parameters $G$ and $R$.


\section{Temperature Dependence of Mutual Information}
\label{sec-tempdep}

Our objective is to compute the mutual information of a region $A \subset \Sigma$ on the left asymptotic boundary and its partner $B\subset \Sigma$ on the right asymptotic boundary, where $B$ is defined so that $A=B$ when the left and right boundaries are identified. In this section we will focus on the simplest case where $\Sigma$ has  zero spatial curvature, $\Sigma = \mathbb{R}^{d-1}$, and the region $A$ is an interval bounded by two hyperplanes. This is so explicit calulations can be performed, though the lessons we learn should extend to other cases. For now we will assume that we have a non-extremal black hole, $f'(R) \neq 0$, and eventually we will be interested in taking the extremal limit where $f'(R)\to 0$. This corresponds to the zero temperature limit, where the mutual information between any pair of partnered left/right regions vanishes.

Let $y$ be a distinguished cartesian coordinate on $\mathbb{R}^{d-1}$, and let $A$ be the region $0 < y < L$. The two boundaries of the region are the hyperplanes $y=0$ and $y=L$.  A minimal bulk surface which shares this boundary is found by extremizing the area functional
\be
\text{Area}_{d-1} = \frac{V_{d-2}}{\lads^{d-2}}\int dr~ r^{d-2}\sqrt{f^{-1} + r^2 y'^2/\lads^2}.
\ee
Here we have used the notation $V_{d-2}$ to denote the volume of a $y={\rm const.}$ hyperplane. Though infinite, we can formally keep track of how it appears in all expressions. Treating this functional as an action, there is a conserved quantity associated with translations in $y$:
\be
\gamma = \frac{r^d}{\sqrt{f^{-1}+r^2y'^2/\lads^2}}y' = r_{\rm min}^{d-1}\lads,
\ee
where $r_{\rm min}$ is the turning point of the surface where $dr/dy = (y')^{-1} = 0$. $r_{\rm min}$ is implicitly a function of $L$, as determined by the constraint
\be
L = \int dy= 2\lads\int_{r_{\rm min}}^{\infty} \frac{dr}{r\sqrt{f}}\, \frac{1}{\sqrt{\left(r/r_{\rm min}\right)^{2d-2}- 1}}.
\ee
The area of the surface is then
\be
\text{Area}_{d-1} = \frac{2V_{d-2}}{\lads^{d-2}}\int_{r_{\rm min}}^\infty dr~\frac{r^{d-2}}{\sqrt{f}}\frac{1}{\sqrt{1-(r_{\rm min}/r)^{2d-2}}}.
\ee
Notice that this area diverges as $r_{\rm min} \to R$. Dividing by $4G$, this gives us the entropy $S(A)$. There is an identical contribution from $S(B)$ on the other side of the geometry. We are left to compute $S(A\cup B)$, which comes from the area of a surface which just passes through the horizon to connect to the other side. By symmetry, this surface is the union of a surface at $y=0$ and a surface at $y=L$. The total area, including both sides of the horizon, is given by
\be
\text{Area}_{d-1} = \frac{4V_{d-2}}{\lads^{d-2}}\int_{R}^\infty dr~\frac{r^{d-2}}{\sqrt{f}}.
\ee
Upon dividing by $4G$ we have $S(A\cup B)$. Putting these results together gives us the mutual information for the interval of width $L$:
\be\label{eq-MI}
I(L) = \frac{V_{d-2}}{G\lads^{d-2}}\left[\int_{r_{\rm min}}^\infty dr~\frac{r^{d-2}}{\sqrt{f}}\frac{1}{\sqrt{1-(r_{\rm min}/r)^{2d-2}}}- \int_{R}^\infty dr~\frac{r^{d-2}}{\sqrt{f}}\right]
\ee
provided the term in brackets is positive, and $I(L)=0$ otherwise. Notice that there is a cancellation of terms over the ranges $r \gg r_{\rm min}$. That is because the $A$ and $A\cup B$ surfaces approximately coincide at large $r$. This can also be seen by noticing that $y' \approx 0$ when $r \gg r_{\rm min}$.

Since we are most interested in where $I(L)$ vanishes, we will approximate Eq.~\ref{eq-MI} in the limit $r_{\rm min} \approx R$. In that case the difference in areas of the two extremal surfaces comes from the difference in area between a segment which hugs the horizon (with area proportional to $L$), and a piece which stretches across the horizon (with area proportional to the proper distance between the horizon and $r_{\rm min}$). Qualitatively, that sort of behavior should persist to situations much more general than the surfaces we are considering. The segment which crosses the horizon will have a proper length which depends on the near-horizon geometry, and especially depends on how close to extremal the black hole is. Examining the non-extremal and near-extremal cases separately, we find that $I(L) \to 0$ when\footnote{The numerical coefficients appearing here are approximate, but the scaling with $f$ and $R$ is exact in the respective limits.}
\be\label{eq-Lcrit}
L \sim \begin{cases}
\frac{2\sqrt{2}\lads}{\sqrt{(d-1)f'(R)R}},~\text{ for } Rf''(R) \ll f'(R),\\
\frac{2\lads}{\sqrt{f''(R)R^2}}\log\left(\frac{Rf''(R)}{f'(R)}\right),~\text{ for } Rf''(R) \gg f'(R).
\end{cases}
\ee
\begin{figure}[t]\hspace*{-.5cm}
\centerline{\includegraphics[width=0.6\columnwidth]{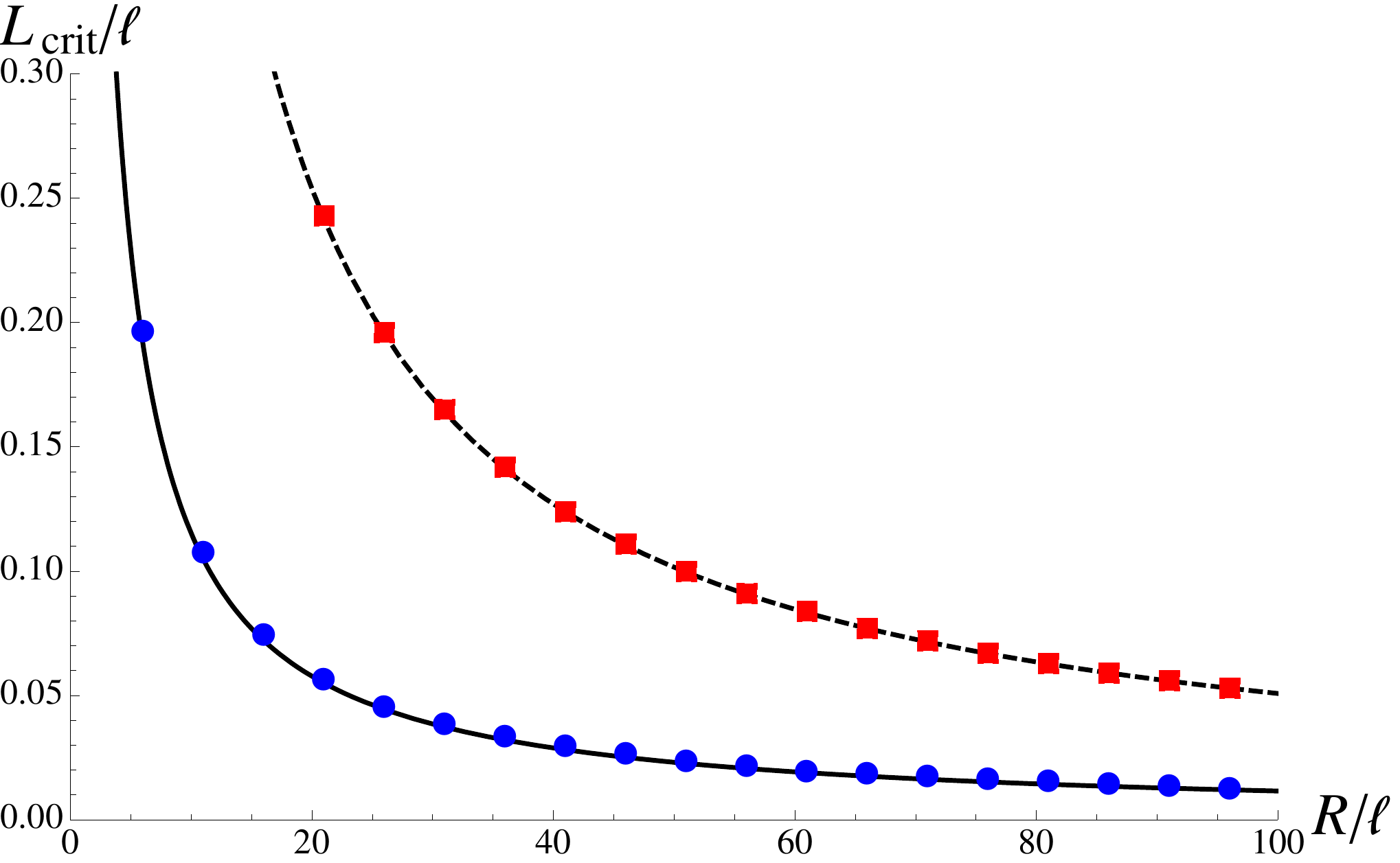}}
\caption{Values of the width $L_{\rm crit}$ of a strip for which the mutual information vanishes. We show results for black holes with different values of $R$ for the uncharged case $q=0$ (blue circles), and the near-extremal case $q=0.999\, q_{\rm ext}$ (red squares). Also shown are the approximations in \Eq{eq-Lcrit}. The non-(near-)extremal approximation is given by the solid (dashed) line. These plots were produced for the specific case $d=3$, $k=1$, but similar results hold for other values of $d$ and $k$.}\label{fig-Lcrit}
\end{figure}
The near-extremal form of this equation was obtained in Ref.~\cite{Andrade:2013rra} for the case of Reissner-Nordstr\"om black holes in AdS$_5$. In \Fig{fig-Lcrit} we plot the value of $L$ for which $I(L)=0$ for several different black holes, as well as the approximation given by \Eq{eq-Lcrit}. Note that for high temperatures, this critical value of $L$ scales like $\beta$. At low temperatures, the critical value of $L$ is controlled by the extremal black hole we are are approaching. It is interesting to note that the logarithmic factor in the near-extremal case can be written as
\be
\log\frac{Rf''(R)}{f'(R)} \sim \log\frac{s}{\Delta s}.
\ee
where $s$ is the entropy density of the zero-temperature extremal black hole and $\Delta s$ is the difference between the near-extremal and extremal entropy densities.

\section{The Butterfly Effect}
\label{sec-butterfly}

\subsection{Shockwave Geometry}

In this section we are insterested in perturbing one side of the geometry by sending in a lightlike pulse of energy fromthe boundary. For simplicity, we consider homogeneous pulses that shift the energy density, $\epsilon \to \epsilon + \delta \epsilon$ with $\delta \epsilon$ small, at fixed $q$.\footnote{Since we are expressing our perturbations in terms of energy and charge, rather than temperature and chemical potential, we are not strictly staying within the grand canonical ensemble. This is not a problem; we know how to add perturbations which carry a fixed amount of charge and energy using AdS/CFT, and the resulting state contains the matter associated with the perturbation in addition to the black hole.} Our first task is to construct the shockwave geometry which accounts for the backreaction of these pulses. We begin with the metric
\be
ds^2 = - f(r) dt^2 + \frac{dr^2}{f(r)} + \frac{r^2}{\lads^2} d\Sigma_{d-1}^2,
\ee
which describes some non-extremal black hole. We define the tortoise coordinate so that $dr_* = dr/f$ and Kruskal coordinates
\be\label{eq-kruskal}
uv = -e^{f'(R) r_*},~~~u/v = -e^{-f'(R) t},
\ee
in terms of which the metric is
\be
ds^2 = \frac{4f(u,v)}{f'(R)^{2}}\frac{dudv}{uv} + \frac{r(u,v)^2}{\lads^2}d\Sigma_{d-1}^2.
\ee
The left exterior region is covered by $u>0$ and the right exterior region by $v >0$. The AdS boundary is located at $uv =  - {\rm const}$, and we can choose this constant to be equal to one by a suitable additive shift in $r_*$ (so $r_*=0$ on the boundary).

\begin{figure}[t]\hspace*{-.5cm}
\centerline{\includegraphics[width=0.9\columnwidth]{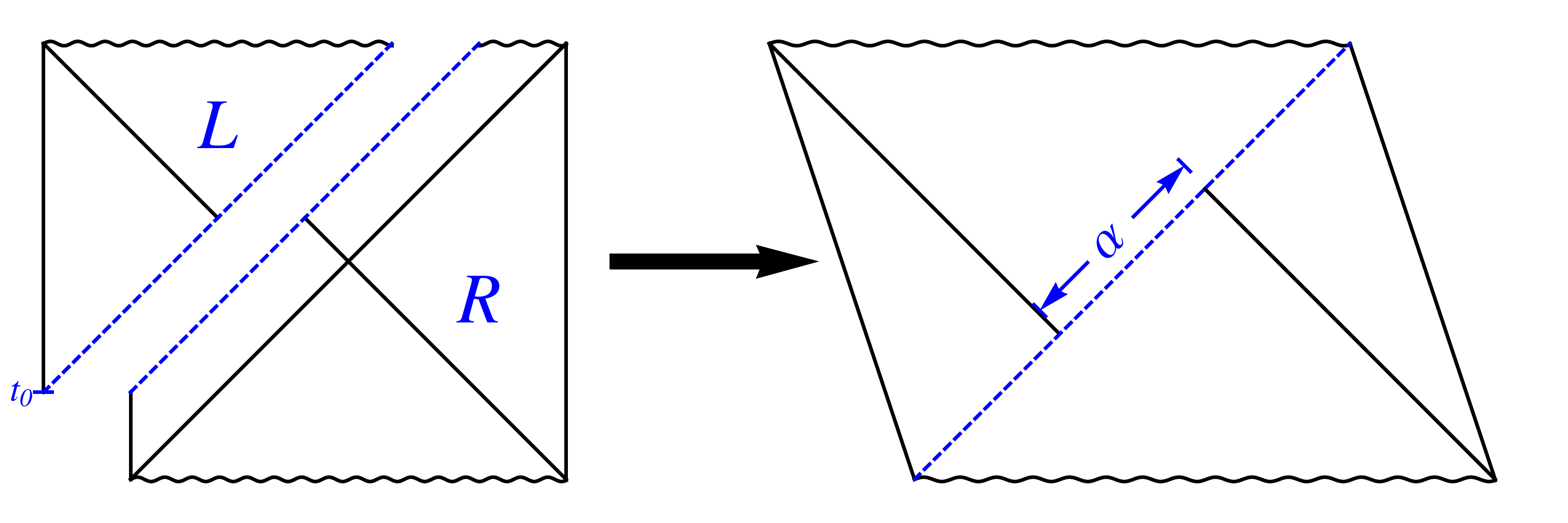}}
\caption{The construction of the shockwave geometry. Two halves ($L$ and $R$) are glued along the lightlike shockwave trajectory. On the right we see the result of the gluing in the $t_0\to \infty$ limit, where the net effect is a shift in the Kruskal coordinate by $\alpha$.}\label{fig-shock}
\end{figure}

A lightlike pulse starting at the left asymptotic boundary follows a constant $u$ trajectory. We will paste together two black hole geometries along this constant $u$ surface. The outside geometry, i.e., in the causal future of the pulse, will be described with coordinates with an $L$ subscript, while coordinates on the inside will have an $R$ subscript (see \Fig{fig-shock}). We can identify $r_L=r_R=r$ because the radius of curvature of the transverse space is continuous across the boundary. Additionally, using time translation symmetry, we can say that the pulse leaves the left asymptotic boundary at $t_R=t_L=t_0$. The pulse is located at $u_{L} = u_{L0}$ and $u_{R} = u_{R0}$ given by 
\begin{align}
u_{L0} &= e^{-2\pi t_0/\beta_L},& u_{R0} &=  e^{-2\pi t_0/\beta_R}.
\end{align}
Since the $r$ coordinate is continuous, we can use the following pair of equaions to find a relationship between the $v_L$ and $v_R$ coordinates along the pulse:
\begin{align}
u_{L0} v_L &= -e^{4\pi r_{L*}(r)/\beta_L},& u_{R0} v_R &= -e^{4\pi r_{R*}(r)/\beta_R}.
\end{align}
We will only consider geometries where $\delta \epsilon$ is small and the time $t_0$ is large. At fixed $v$, large $t_0$ means taking $u$ to zero, which in turn means $r_*$ approaches $-\infty$. In other words, we approach the horizon. Near the horizon of a non-extremal black hole, $r_* \approx \frac{\beta}{4\pi}\left(C + \log(r-R)\right)$, where $C$ is a constant that  depends on the geometry. Hence we have
\begin{align}
u_{L0} v_L &\approx -(r-R_L)e^{C_L},& u_{R0} v_R &\approx -(r-R_R)e^{C_R}.
\end{align}
In the limit that $R_L-R_R \equiv \delta R \to 0$ while $e^{2\pi t_0/\beta_R}\delta R$ remains finite, we have $\beta_R = \beta_L = \beta$, $C_R = C_L = C$, the Kruskal coordinate $u$ becomes continuous (and hence we will drop subscripts on it), while along $u=0$ we have the identification
\be
v_L = v_R + e^Ce^{2\pi t_0/\beta}\delta R \equiv v_R + \alpha.
\ee


Thus the matching condition is given by a shift in the Kruskal coordinate $v$. This leaves both the left exterior and the right exterior individually unaffected, but any spacelike probe which crosses from left to right will feel the influence. In particular, for the probes we consider, we should expect large deviations compared to the shockwave-less result when $\alpha\sim 1$, or in other words when $t_0 \sim t_*$ where
\be
t_* \equiv \frac{\beta}{2\pi}\log\frac{e^{-C}}{\delta R}.
\ee
We will see this sort of behavior in examples below. A remaining challenge is to find an expression for $C$. While in simple cases analytic expressions can be found, there are two interesting limits to consider: the high temperature limit and the low temperature limit at fixed $q$.

The high temperature limit should be represented by an uncharged planar black hole, for which we can compute
\be
r_*(r)  = -\lads^2 \int_r^\infty\frac{dr'}{r'^2 - \frac{R^d}{r'^{d-2}}} \approx -\frac{\beta}{4\pi}\left[\log d + \log\frac{r-R}{R} + \psi(1/d) + \gamma\right]+ O(r-R),
\ee
where $\psi$ is the digamma function and $\gamma$ is the Euler-Mascheroni constant. The resulting expression for $t_*$ is
\be
t_* = \frac{\beta}{2\pi}\log \frac{e^{-\psi(1/d)-\gamma}R}{d\delta R} =  \frac{\beta}{2\pi}\log \frac{e^{-\psi(1/d)-\gamma}\epsilon}{\delta \epsilon}.
\ee
This matches the high-temperature limit of the exact calculation in \App{sec-exact} for a black holes in $d=4$, as it should. The energy-independent factor in the logarithm technically only gives a subleading contribution to $t_*$, but we include it anyway for the purposes of comparison to other similarly precise calculations.

The low temperature limit corresponds to a near-extemal black hole. As we show in \App{sec-nearext}, a near-extremal black hole satisfies
\be
t_* = \frac{\beta}{2\pi} \log \frac{4\Delta \epsilon}{\delta \epsilon}.
\ee
Here we have used the notation $\Delta \epsilon = \epsilon - \epsilon_{\rm ext}$, the energy density in excess of the extremal energy density. The universality of this expression is remarkable, as it holds for charged extremal black holes of any charge as well as uncharged, hyperbolic black holes. It is interesting that we are required to use $\Delta \epsilon$ rather than $\epsilon$ itself. As we discussed in the introduction, this suggests a division of degrees of freedom into the extremal degrees of freedom and the excited degrees of freedom. Here we see that only the excited degrees of freedom contribute to the butterfly effect. In the previous section, we saw that the extremal degrees of freedom did not contribute to the mutual information as we lower the temperature. We also note that, once again, this result agrees with the zero-temperature limit of the exact calculations in \App{sec-exact} for $d=4$.

Both the high temperature and low temperature limits are readily expressed in terms of $\log \epsilon/\delta \epsilon$, as long as we understand ``$\epsilon$" to mean energy above the zero temperature state, and we see that there is a temperature-dependent subleading piece which also depends on the spatial dimension. We conjecture that the leading term of $t_*$ is always given by $\frac{\beta}{2\pi} \log \epsilon/\delta\epsilon$, though we are unable to prove it.

\subsection{Extremal Surfaces}

While we expect the parameter $\alpha$ of the shockwave geometry to control the disruption of mutual information, it is useful to show explicitly that this is the case at least in some simple examples. As a probe of the butterfly effect we wish to compute the mutual information of a region $A$ on the left asymptotic boundary and its identical partner $B$ on the right asymptotic boundary, at $t=0$. The surfaces representing $S(A)$ and $S(B)$ are unaffected by the shockwave because they do not cross the horizon. However, the surface with boundary $A\cup B$ which stretches across the wormhole is disrupted by the shockwave, and our task in this section is to compute its new shape and area as a function of $\alpha$.

We will consider the case where $A \subset \Sigma$ is half of the space. By symmetry, the minimal surface in the bulk will always divide the transverse space in half, so the problem of finding the minimal surface is reduced to a two dimensional problem. The area of the minimal surface is then given by
\be
\text{Area}_{d-1} =  \frac{V_{d-2}}{\lads^{d-2}} \int dt~r^{d-2}\sqrt{-f  +f^{-1}\dot{r}^2}.
\ee
Here we are using $V_{d-2}$ to denote the volume of the lower-dimensional surface which divides the transverse space in half (i.e., a $(d-2)$-plane in the $k=0$ case or the equator in the $k=1$ case) computed using the $d\Sigma_{d-1}^2$ line element. Treating the area functional as a single-particle action, the conserved quantity associated to $t$-translation is
\be
\gamma = \frac{-f r^{d-2}}{\sqrt{-f + f^{-1}\dot{r}^2}} =  \sqrt{-f_0}r_0^{d-2}.
\ee
Here we have defined $r_0$ as the radial position where $\dot{r}=0$, and $f_0= f(r_0)$, which is presumed negative since this point will be behind the (outer) horizon. In the limit that $r_0\to R$ we have $\gamma \to 0$, and this should correspond to the limit $\alpha \to 0$ where the shockwave is absent.

For future reference, we note that the $t$ coordinate as a function of the radius is given by
\be\label{eq-time}
t(r) = \int \frac{dr}{f\sqrt{1 + \gamma^{-2}fr^{2d-4}}}.
\ee
Our main objective is to compute the area of the portion of the minimal surface that starts at $t=0$ on the left asymptotic boundary and ends at $v=\alpha/2$ on the horizon (suppressing the $L$ subscript on $v$). We can double the area of this half of the surface to get the total area.

\subsubsection{Surface Location}

\begin{figure}[t]\hspace*{-.5cm}
\centerline{\includegraphics[width=0.8\columnwidth]{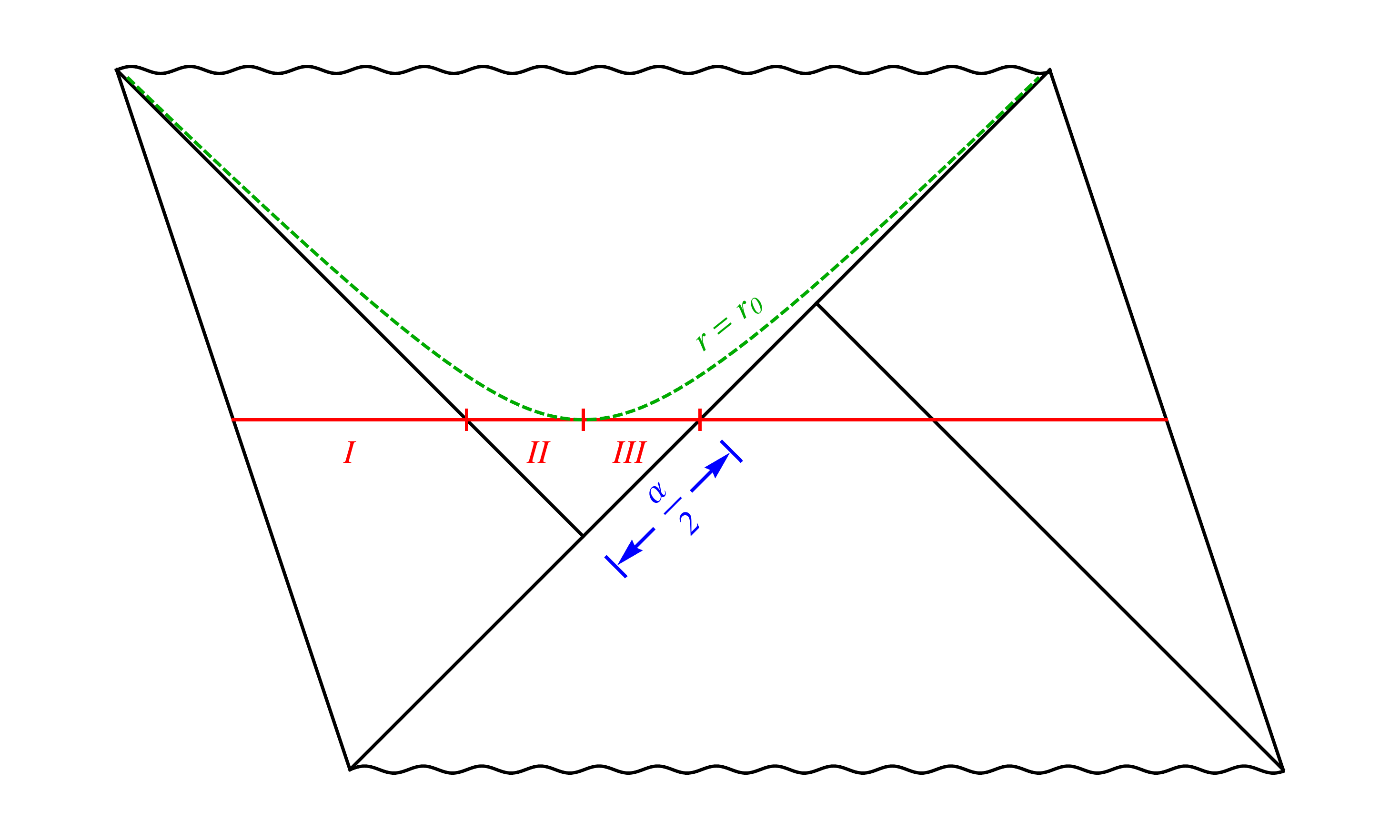}}
\caption{The minimal surface (horizontal, red) in the shockwave geometry. We split the left half of the surface into three segments, labeled I, II, and III in the figure, to aid in calculation. The division between I and II occurs at the left future horizon. The smallest value of $r$ attained by the surface is $r=r_0$, which marks the division between II and III.}\label{fig-surf}
\end{figure}

 It will be useful to find a relationship between $\alpha$ and $\gamma$. We can do this by splitting the left half of the surface into three segments, as in \Fig{fig-surf}. The first segment goes from the boundary to $v=0$ (at some value of $u$), the second from $v=0$ to $r = r_0$ (at some value of $t$), and the third from $r=r_0$ to $u=0$. 

For the first segment, the minimal surface stretches from the boundary at $(u,v) = (1,-1)$ to $(u,v) = (u_1,0)$. Using \Eq{eq-kruskal} we have
\be
u_1^2 = \exp\left[f'(R)(\Delta r_* - \Delta t)\right] = \exp\left[-\frac{4\pi}{\beta}\int_R^\infty \frac{dr}{f}\left(1-\frac{1}{\sqrt{1+ \gamma^{-2}f r^{2d-4}}}\right) \right].
\ee
For the second segment, the minimal surface stretches from $(u,v) = (u_1,0)$ to $(u,v) = (u_2,v_2)$. We know that $(u_2,v_2)$ lies on the surface $r=r_0$ but we don't know at what value of $t$. We do know that
\be
\frac{u_2^2}{u_1^2} = \exp\left[-\frac{4\pi}{\beta}\int_{r_0}^R \frac{dr}{-f}\left(\frac{1}{\sqrt{1+ \gamma^{-2}f r^{2d-4}}}-1\right) \right].
\ee
So we have
\be
u_2^2 =  \exp\left[-\frac{4\pi}{\beta}\int_{r_0}^\infty \frac{dr}{f}\left(1-\frac{1}{\sqrt{1+ \gamma^{-2}f r^{2d-4}}}\right) \right].
\ee
To find $v_2$, it is simplest to use a reference surface $r=\bar{r}$ for which $r_*=0$ in the black hole interior.\footnote{If there is no such surface lying away from the singularity, we can use a different reference surface. The precise value of $r_*$ where $r=\bar{r}$ does not matter.} Then
\be
v_2 = \frac{1}{u_2}\exp\left( -\frac{4\pi}{\beta} \int_{\bar{r}}^{r_0}\frac{dr}{-f}\right).
\ee
For the final segment, we have
\be
\frac{\alpha^2}{4v_2^2} = \exp\left[f'(R)\left(\Delta r_* + \Delta t\right)\right] = \exp\left[\frac{4\pi}{\beta}\int_{r_0}^R \frac{dr}{-f}\left(\frac{1}{\sqrt{1+ \gamma^{-2}f r^{2d-4}}}-1\right) \right] = \frac{u_1^2}{u_2^2}.
\ee
Putting all of this together we have
\be
\alpha = 2 \exp(K_1+K_2+K_3)
\ee
where
\begin{align}
K_1 &= -\frac{4\pi}{\beta} \int_{\bar{r}}^{r_0}\frac{dr}{-f}\,,\\
K_2 &=\frac{2\pi}{\beta}\int_{R}^\infty \frac{dr}{f}\left(1-\frac{1}{\sqrt{1+ \gamma^{-2}f r^{2d-4}}}\right),\\
K_3 &= \frac{4\pi}{\beta}\int_{r_0}^R \frac{dr}{-f}\left(\frac{1}{\sqrt{1+ \gamma^{-2}f r^{2d-4}}}-1\right).
\end{align}
This is an equation that relates $r_0$ to $\alpha$. We would like to identify the limits of $r_0$ in which $\alpha$ vanishes and in which $\alpha$ becomes large. Either of these behaviors requires one or more of $K_1$, $K_2$, and $K_3$ to diverge. $K_1$ and $K_2$ diverge as $r_0\to R$, and (it turns out) this corresponds to $\alpha \to 0$. $K_3$ diverges when $r_0\to r_{\rm crit}$ where
\be
f'(r_{\rm crit})r_{\rm crit} +(2d-4)f(r_{\rm crit}) = 0.
\ee
This limit corresponts to $\alpha \to \infty$.

Let us first dispense with the case $r_0 / R \approx 1$, which means that the surface does not go far behind the horizon. The first integral diverges logarithmically in the usual way for the tortoise coordinate:
\be
K_1 \approx \log(R- r_0) + \ldots
\ee
The $K_2$ integral is similar. The factor in parantheses in the $K_2$ integrand cuts off the logarithmically diverging $1/f$ factor at the value of $r$ for which $fr^{2d-4} = -f(r_0)r_0^{2d-4}$. When $r_0 \approx R$ this corresponds to cutting off the integration at $r\approx 2R-r_0$. Then we have
\be
K_2 \approx -\frac{1}{2} \log(R-r_0) +\ldots
\ee
The integral $K_3$ remains finite. Thus we see that $\alpha \propto \sqrt{R-r_0}$ as $R\to r_0$.

The other interesting limit is when $r_0\approx r_{\rm crit}$. Here $K_1$ and $K_2$ approach finite values, while $K_3$ diverges. This divergence is logarithmic as $r_0\to r_{\rm crit}$:
\be
K_3 \propto \log\frac{1}{f'(r_0)r_0+(2d-4)f(r_0)} + \cdots
\ee
We see that $\Delta t$ diverges as well in this limit using \Eq{eq-time}. That means the minimal surface is hugging the surface $r = r_{\rm crit}$ for a large interval of time. This behavior was also discussed in Ref.~\cite{Hartman:2013qma}.

\subsubsection{Surface Area}

The area of the minimal surface is given by
\be
\text{Area}_{d-1} = \frac{V_{d-2}}{\lads^{d-2}} \int dr~r^{d-2}\frac{1}{\sqrt{\gamma^2r^{4-2d}+f}}.
\ee
We can compute the area for each segment define above, and multiply the answer by two to get the total area. The second and third segments ($R$ to $r_0$ and back to $R$) manifestly have the same area. Hence the total area is given by the twice the area of the first segment plus four times the area of the second segment. The first segment contains a divergent $\alpha$-independent contribution which must be subtracted.

The $\alpha$-dependent part of the area at large $\alpha$ comes from the long part of the surface near $r\approx r_{\rm crit}$. In that limit, the area functional becomes proportional to the $K_3$ integral we encountered above, and so we find
\be
\text{Area}_{d-1} \approx \frac{4V_{d-2}}{\lads^{d-2}} \frac{\beta}{4\pi} r_{\rm crit}^{d-2}\sqrt{-f_{\rm crit}}\log \alpha +\cdots.
\ee
As promised, the area increases as a function of $\alpha$. The area depends on $\alpha$ logarithmically, which means it depends linearly on the shockwave time $t_0$.

\section{Discussion}

We have examined in detail two ways in which the mutual information between local regions of two entangled CFTs may be disrupted. On the AdS side, the mutual information between localized regions is roughly dual to the length of the wormhole. Disrupting the mutual information leads to a long wormhole, and in the limit of an infinitely long wormhole we can interpret the spacetimes as being disconnected. This is an intermediate case between the classical wormhole and the fully ``quantum" wormhole of Maldacena and Susskind, which does not have any classically geometric properties.

The first way in which the mutual information is disrupted is simply through lowering the temperature. At zero temperature, an extremal black hole has an infinitely long wormhole, and we consider it an open question whether or not this should be considered a classial wormhole. If we examine the extremal solution directly, then a two-sided extremal black hole actually consists of two disconnected spacetimes.  Then it is difficult to regard the interior geometry as arising from the entanglement of the two exterior regions, as one does for a non-extremal black hole. The situation may be more like the Poincare patch of vacuum AdS, where the region behind the horizon exists but does not arise from degrees of freedom in a second, independent, CFT.\footnote{We can consider stacking two Poincare patches on top of each other and say that the CFT in the boundary of one patch describes the physics behind the horizon of the other patch. However, in that case the two CFTs can be related through conjugation by rotation and time translation in the global picture.} However, we note that the interior of an extremal black hole is very subtle, as the extremal limit of near-extremal black holes does not coincide with the extremal black hole itself~\cite{Carroll:2009maa}. One should regard the extremal limit of near-extremal black holes as the physical solution, and we leave a careful analysis of its properties for future work.


Our second method of disrupting mutual information was to make use of the butterfly effect Ref.~\cite{Shenker:2013pqa}. A single shockwave leads to a lengthening of the wormhole connecting the two exterior regions after a time $t_* \sim \frac{\beta}{2\pi} \log E/\delta E$. Multiple shockwaves will add to this effect. The analysis of Ref.~\cite{Shenker:2013yza} for multiple shocks can be carried out again in our more general setting, leading to very long wormholes. An open question is how well these multiple shock geometries capture the properties of a typical state in the ensemble. It may be the case that a typical state contains a smooth wormhole, albeit a very long one. However, another possibility is that a typical state contains a firewall behind the horizon, with no classical wormhole connecting the two sides.

We conclude by commenting on the butterfly effect in the vacuum.\footnote{I thank Stephen Shenker and Douglas Stanford for a discussion of this point.} Notice that the special case $k=-1$, $\mu=0$ of our setup is the hyperbolic slicing of empty AdS. In this case the two-sided ``black hole" is just the Rindler decomposition of the AdS vacuum, analogous to what is usually done in flat space. Our analysis of the butterfly effect here seems to show that chaotic dynamics is important even for the vacuum state. This statement is a bit strange, given that the near-vacuum dynamics of ${\mathcal N}=4$ Super Yang-Mills theory at large $N$ is integrable, for example. There are two caveats that need to be mentioned which explain this behavior. First is that our perturbations are homogeneous in the hyperbolic slicing, which means that they have infinite energy in the global slicing. This can be alleviated by considering finite-volume perturbations in the hyperbolic slicing, and even though this makes the analysis technically more challenging, there is no reason to expect the conclusions to drastically change. Second, and most importantly, is that the shockwave limit $t_0\to \infty$ in the hyperbolic slicing represents an infinite boost in the global slicing. In other words, a small perturbation of Rindler energy $\delta E$ released at $t_0$ has global energy $\delta E \exp(t_0)$, and for $t_0\sim t_*$ this corresponds to an energy of order $N^2$, enough to make a big black hole.

\begin{center}{\bf \large Acknowledgements}\end{center}
 
I would like to thank Sean Carroll, Steve Shenker, Douglas Stanford, and Lenny Susskind for helpful comments and discussions.   This research is supported by the DOE under Contract No.~DE-SC0011632 and the Gordon and Betty Moore Foundation through Grant No.~776 to the Caltech Moore Center for Theoretical Cosmology and Physics, as well as a John A. McCone Postdoctoral Fellowship.  I would also like to thank the Aspen Center for Physics and the participants of the New Perspectives on Thermalization conference where this work was initiated.

\appendix

\section{RNAdS Thermodynamics}
\label{sec-RNthermo}

In this appendix we review the thermodynamics of Reissner-Nordstr\"om AdS black holes~\cite{Peca:1998cs, Chamblin:1999tk, Hawking:1999dp}. The metric takes the form
\be
-f(r) dt^2 + f(r)^{-1} dr^2 + \frac{r^2}{\lads^2}d\Sigma_{d-1}^2~,
\ee
where $\Sigma$ is a sphere $(k=1)$, a hyperboloid\footnote{In the $k=\pm 1$ cases, the radius of curvature of the transverse space is $\lads$.} $(k=-1)$, or a plane $(k=0)$, and the metric factor is
\be
f(r) = k - \frac{\mu}{r^{d-2}}+\frac{q^2}{r^{2d-4}} + \frac{r^2}{\lads^2}.
\ee
The electric potential is given by
\be
A_t = \frac{q}{\sqrt{8\pi G}}\sqrt{\frac{d-1}{d-2}}\left(\frac{1}{r^{d-2}} - \frac{1}{R^{d-2}}\right) \equiv \frac{1}{\sqrt{8\pi G}}\sqrt{\frac{d-1}{d-2}}\left(\frac{q}{r^{d-2}} +\phi \right).
\ee
The chemical potential $\phi$ is the (appropriately rescaled) potential difference between the horizon and infinity. The temperature, as usual, is determined by requiring that the Euclidean geometry be nonsingular at $r=R$:
\be
\frac{4\pi}{\beta} = f'(R) = d\frac{R}{\lads^2} +(d-2)\frac{k}{R} - (d-2)\frac{q^2}{R^{2d-3}}= d\frac{R}{\lads^2} +(d-2)\frac{k-\phi^2}{R}.
\ee
The free energy is computed by calculating the regularized Euclidean action,
\begin{align}
S_E(\beta, \phi) &= -\frac{1}{16\pi G}\int d^{d+1}x \sqrt{g}\left(\mathcal{R} + \frac{d(d-1)}{\ell^2}\right) - \frac{1}{4}\int d^{d+1}x\sqrt{g} F_{\mu\nu}F^{\mu\nu} - S_0\\
&=\frac{\beta}{16\pi G }\left[(k-\phi^2)R^{d-2}  - \frac{R^d}{\lads^2} +\delta_{k,-1}2\lads^{d-2}\frac{d-1}{d-2}\left(\frac{d-2}{d}\right)^{d/2}\right].
\end{align}
We should think of this action as being a function of the thermodynamic variables $\beta$ and $\phi$ which are conjugate to the energy and charge. The subtraction $S_0$ is made to render the action finite, and is independent of the thermodynamic variables, and sets the zero of the free energy. For $k=1,0$ we use vacuum AdS as our zero point, but for $k=-1$ we cannot. This is because the zero point must be a zero-temperature solution in order to perform the subtraction, since for technical reasons we require that the Euclidean time be allowed to have any period. For $k=-1$ we will choose the solution with $q=0$ and $\mu = \mu_0 \equiv  -\frac{2\lads^{d-2}}{d-2}\left(\frac{d-2}{d}\right)^{d/2}$ as our zero-point.

We can compute the thermal averages for charge and energy by taking derivatives:
\begin{align}
\langle Q \rangle &\equiv \frac{1}{\beta}\frac{\partial S_E}{\partial \phi} = \frac{(d-1){\rm Vol}(\Sigma)}{8\pi G \lads^{d-1}} q, & \langle E \rangle&\equiv \frac{\partial S_E}{\partial \beta} - \phi \langle Q \rangle = \frac{(d-1){\rm Vol}(\Sigma)}{16\pi G \lads^{d-1}}(\mu-\mu_0).
\end{align}
The entropy is given by the Legendre transform of the Euclidean action with respect to $\beta$, and coincides with the horizon area divided by $4G$:
\be
S \equiv \beta\frac{\partial S_E}{\partial \beta} - S_E = \frac{1}{4G}\left(\frac{R}{\lads}\right)^{d-1}{\rm Vol}(\Sigma).
\ee
It's also useful to make note of the First Law of Thermodynamics in terms of the black hole paramters $\mu$, $q$, and $R$:
\be
0 = \beta \delta \mu +2 \beta \phi \delta q - 4\pi R^{d-2} \delta R~.
\ee
\section{Exact Results in $d=4$}
\label{sec-exact}

In this appendix we note closed-form expressions for the tortoise coordinate and $t_*$ at all values of the temperature for black holes in $d=4$. The metric is
\be
ds^2 = -\left(\frac{r^2}{\ell^2} -\frac{\mu}{r^2} + k\right) dt^2 + \frac{1}{\frac{r^2}{\ell^2} -\frac{\mu}{r^2} + k}dr^2 + \frac{r^2}{\lads^2}d\Sigma_2^2,
\ee
and we will consider only the cases $k=\pm1$. The $k=0$ case can be obtained as a limit of the others. It is also convenient to treat the case $\mu>0$ separately from $\mu<0$ (which is possible only when $k=-1$). For $\mu>0$, define
\be
\frac{R_\pm^2}{\lads^2} = \pm \frac{-k}{2} + \frac{1}{2}\sqrt{k^2 + 4\mu\lads^{-2} }.
\ee
Here $R_+ = R$ is the horizon, and $R_-$ is just an auxilliary variable. Then we have
\begin{align}
r_*(r) &= -\int_r^\infty \frac{dr'}{k - \frac{\mu}{r'^2} + \frac{r'^2}{\lads^2}} =  -\int_r^\infty \frac{\lads^2r'^2 dr'}{(r'^2 - R_+^2)(r'^2 + R_-^2)} \\
&=-\frac{\beta}{4\pi R_+} \left[2R_- \tan^{-1}\left(\frac{R_-}{r}\right)+R_+ \log(\frac{r+R_+}{r-R_+})\right].
\end{align}
From here we can extract
\be
C(\beta) = -2\frac{R_-}{R_+}\tan^{-1}\frac{R_-}{R_+} - \log 2R_+,
\ee
from which we can compute $t_*$. In the large $\mu$ limit, which corresponds to large temperatures, we find $C(\beta) =  -\frac{\pi}{2} - \log 2R$ and we have
\be
t_* = \frac{\beta}{2\pi}\log \frac{2e^{\pi/2}R}{\delta R} = \frac{\beta}{2\pi}\log \frac{8e^{\pi/2}\epsilon}{\delta \epsilon}.
\ee
The numerical factor within the logarithm is an unimportant subleading factor, but it's useful to keep around both to facilitate detailed comparisons to other limiting cases and to confirm that the answer matches the exact calculation of the planar $k=0$ case, which it does.

Finally we turn to $\mu<0$, which is only valid when $k=-1$. Now we define
\be
R_{\pm}^2/\lads^2  = \frac{1}{2} \pm \frac{1}{2}\sqrt{1 + 4 \mu \lads^{-2}},
\ee
and here $R_-$ is real and corresponds to the inner horizon. The tortoise coordinate can again be computed exactly:
\begin{align}
r_* &= \int_r^\infty \frac{dr}{-1 - \frac{\mu}{r^2} + \frac{r^2}{\lads^2}} =  \int_r^\infty \frac{\lads^2r^2 dr}{(r^2 - R_+^2)(r^2 - R_-^2)} \\
&=\frac{\beta}{4\pi R_+} \left[R_- \log(\frac{r+R_-}{r-R_-})-R_+ \log(\frac{r+R_+}{r-R_+})\right].
\end{align}
The function $C(\beta)$ can be extracted:
\be
C(\beta) = \frac{R_-}{R_+}\log(\frac{R_+ + R_-}{R_+-R_-}) - \log 2R_+,
\ee
which lets us compute $t_*$. The interesting limit now is the low temperature limit where $\mu \lads^{-2} \approx -1/4$. In that case we find $C(\beta) = -\log(\pi\lads^2/\beta)$, and so we get
\be
t_* = \frac{\beta}{2\pi}\log \frac{\pi \lads^2}{\beta \delta R} = \frac{\beta}{2\pi}\log \frac{4 \epsilon }{\delta \epsilon}.
\ee

\section{Near-Extremal Black Holes}
\label{sec-nearext}

In this appendix we will compute the time $t_*$ for a near-extremal black hole. A near-extremal black hole has $f = f_{\rm ext} - \frac{\Delta \mu}{r^{d-2}}$ where $\Delta\mu$ is small and $f_{\rm ext}$ has a double zero at $r=R_{\rm ext}$.\footnote{There are multiple ways to deform an extremal black hole to a near-extremal one. We always consider fixed charge in this work, which we find convenient.} The horizon is at $r=R = R_{\rm ext}(1+ \kappa)$. To lowest order in $\Delta \mu$, $\kappa$ is given by
\be
\kappa \approx \sqrt{\frac{2\Delta \mu}{f_{\rm ext}''(R_{\rm ext})R_{\rm ext}^d}} \approx \sqrt{\frac{2\Delta \mu}{f''(R)R^d}}.
\ee
The temperature of this black hole is computed from
\be
\frac{4\pi}{\beta} = f'(R) = f_{\rm ext}'(R) +(d-2)\frac{\Delta \mu}{R^{d-3}} \approx f_{\rm ext}''(R_{\rm ext})R_{\rm ext} \kappa  \approx \sqrt{\frac{2f''(R)\Delta \mu}{R^{d-2}}}.
\ee
In the near-extremal limit we can compute the total change in tortoise coordinate from the boundary to the horizon. This consists of a piece that diverges as we approach the horizon and a finite piece that diverges as we take $\Delta\mu\to 0$. Letting $\lambda$ be some small constant, we have
\begin{align}
r_* &= -\int_r^\infty \frac{dr}{f} =  {\rm const.} -\int_{r-R}^{\lambda R} \frac{dx}{f'(R)x + \frac{1}{2}f''(R)x^2} \\
&\approx  {\rm const.} +\frac{1}{\lambda f''(R) R}+ \frac{1}{f'(R)}\left(\log \frac{f''(R)R}{f'(R)} + \log\frac{r-R}{2R} \right).
\end{align}
Using this result, we find that
\be
t_* = \frac{\beta}{2\pi}\log \frac{2f'(R)}{f''(R) \delta R}=  \frac{\beta}{2\pi}\log \frac{4 \Delta\epsilon }{\delta \epsilon},
\ee
where $\Delta \epsilon = \epsilon-\epsilon_{\rm ext}$ is the energy density in excess of the extremal black hole we are approximating. We are implicitly assuming that $\delta \epsilon \ll \Delta \epsilon$. Note that for hyperbolic black holes of zero charge, as considered in \App{sec-exact}, $\Delta \epsilon = \epsilon$ because of how we chose the zero point of the energy.

\bibliographystyle{utcaps}
\bibliography{BHentanglement}

\end{document}